\documentclass[doublecol]{epl2} 
\title{The Goldstone theorem  in nonlinear electrodynamics}
\author{C. A. Escobar\inst{1} \and L. F. Urrutia\inst{1,2} }
\shortauthor{C. A. Escobar \etal }
\institute{
  \inst{1} Instituto de Ciencias Nucleares, Universidad Nacional Aut{\'o}noma de
M{\'e}xico,
A. Postal 70-543, 04510 M{\'e}xico D.F., M{\'e}xico\\
  \inst{2}Facultad de F\'{\i}sica, Pontificia Universidad Cat\'{o}lica de
Chile, Casilla 306, Santiago 22, Chile}
\pacs{11.30.Qc}{Spontaneous and radiative symmetry breaking}
\pacs{11.30.Cp}{Lorentz and Poincare invariance}
\pacs{11.10.Lm}{Nonlinear or nonlocal theories and models}
\abstract{
We construct a non-linear electrodynamics arising from the spontaneous
Lorentz symmetry breaking triggered by a non-zero vacuum expectation value of
the electromagnetic field strength, instead of the electromagnetic potential.
The expansion of the corresponding
action in terms of the physical photon excitations is naturally written in
terms of generalized permeabilities, in a manner similar to when the polarization is expanded in terms of
the susceptibilities when studying non-linear effects in standard optics.
The new version of the Goldstone theorem is investigated and implies that the lowest order
permeability, been the analogous of the mass matrix in the standard case,
has zero modes. These are interpreted as the corresponding Goldstone modes
of the model and they are explicitly constructed for one of the two choices of
vacua that the theory admits. Further iterations of the method employed to
obtain such theorem yield  relationships among the generalized
permeabilities. Some preliminary results regarding the splitting of the two degrees of freedom of the model into
Goldstone and non-Goldstone modes are presented.}

\begin{document}
\maketitle

\section{(I) Introduction}
Spontaneously Lorentz symmetry breaking (SLSB) has played an important role
in the construction of models  designed to probe the validity of such
symmetry. Indeed, the whole Standard Model Extension (SME) \cite{KOSTELECKY1,KOSTELECKY2},
which is one of the most used frameworks to parameterize  Lorentz-invariance-violating phenomena, is assumed to arise as a consequence of \ SLSB
occurring in a more fundamental theory \cite{KOSTELECKY3}.
Moreover, in the gravity sector of the SME it has been proved that explicit
breaking is inconsistent with the geometry of the curved manifold, encoded
in the Bianchi identities \cite{KOSTELECKY4}. On the other hand SLSB has
also been studied  as a mechanism to provide a dynamical setting for the
gauge principle in field theory \cite{GBPHOTONS1}. The basic idea is that gauge particles, such as
photons, gluons and gravitons would be realized as the Goldstone bosons \
arising via SLSB from a non-gauge invariant theory \cite{GBPHOTONS2}.

SLSB requires that non-zero vacuum expectation values (VEV) are generated
from tensor fields in such a way that preferred directions are selected in
 space-time. Most of the cases discussed in the literature consider
the Goldstone mode excitations to be associated with the zero eigenvalues of
the mass matrix of the problem, as a consequence of the standard
interpretation of the Goldstone theorem. This is also  the case when SLSB
caused by VEV of antisymmetric tensor fields $B_{\mu \nu }$ is considered
\cite{KOSTELECKY5}. Here $B_{\mu \nu }$ is taken to be  the potential
for a field strength $H_{\alpha \beta \gamma }$, (i. e. $H={\rm d}B$), and its mass
matrix is investigated.
In this work we further study spontaneous Lorentz symmetry breaking in an
effective non-linear electrodynamics (NLED), triggered by a non-zero vacuum expectation value $C_{\mu\nu}$ of the
field strength $F_{\mu\nu}$, which is provided by  the minimum of a potential $V(F_{\mu\nu})$
arising from a fundamental theory. This model was originally proposed in Ref. \cite{JALU}, where
some  particular limits leading to Lorentz invariance violating (LIV) modifications of standard electrodynamics were discussed. Since the
theory is gauge invariant, all the excitations are massless, and an
alternative interpretation of the Goldstone theorem is required. The resulting
Goldstone matrix, exhibiting  zero modes, is no longer a mass matrix but rather a permeability matrix.
In this letter we  provide a
discussion of the corresponding Goldstone theorem, including a general
construction of the associated Goldstone modes (GM), together with the properties and characterization of the related model.
\section{(II) The equations in non-linear electrodynamics}
We start from a
formulation of NLED which is alternative to that of Ref.\cite{PLEBANSKI}. It   has  the  advantage of not requiring to invert the non-linear relations that define the fields
$\mathbf{E}, \mathbf{B}$ in terms of $\mathbf{D}, \mathbf{H}$.
The price one has to pay is that the coupling to the electromagnetic current in the unbroken action
appears in a non-local way, since we insist that our initial variable is the
field strength $F_{\alpha \beta }$, instead of the potential $A_{\nu }$.
Our conventions are the same as in Ref.\cite{JACKSON}.
The action is \cite{JALU},
\begin{equation}
S=\frac{1}{2}\int d^{4}x\;\left( - V(F_{\alpha \beta })-%
\epsilon ^{\nu \mu \alpha \beta }F_{\alpha \beta }\partial _{\nu
}X_{\mu } +  F_{\mu \nu }J^{\mu \nu }\right),\label{ACT1}
\end{equation}
with
\begin{equation}
 J^{\mu \nu }=\int d^{4}x^{\prime }\;\left[ f^{\mu
}(x-x^{\prime })J^{\nu }(x^{\prime })-f^{\nu }(x-x^{\prime })J^{\mu
}(x^{\prime })\right]. \label{CURRENT}
\end{equation}%
which  is invariant under the gauge
transformations
$
X_{\mu }\rightarrow X_{\mu }+\partial _{\mu }X.  \label{GT}
$
The distribution $f^{\mu }(x-x^{\prime })$ was introduced by Schwinger in
his discussion of magnetic charge, together with the description of
different models of electromagnetic sources in QED  and
satisfies
$
\partial _{\mu }f^{\mu }(x-x^{\prime })=\delta ^{(4)}(x-x^{\prime })
$ \cite{SCHW}.
The equations of motion arising from the action (\ref{ACT1}) are
\begin{eqnarray}
\epsilon ^{\nu \mu \alpha \beta
}\partial _{\nu }F_{\alpha \beta }=0, \qquad  0=\frac{\partial V}{\partial
F_{\alpha \beta }}+\epsilon ^{\nu \mu \alpha \beta }\partial _{\nu }X_{\mu
}-J^{\alpha \beta }.  \label{EQ4}
\end{eqnarray}%
 We have explicitly verified that the action (\ref{ACT1}) describes two degrees of freedom (DOF) in coordinate space.
The fields $X_{\mu }$ play the role of  Lagrange multipliers, which  lead to the existence of the potential $%
A^{\nu }$ for the field strength, as a consequence of  the Bianchi identity provided by the first equation in (\ref{EQ4}). The Lagrange multipliers can be eliminated by taking the divergence of the second equation in (\ref{EQ4}),
leading to the non-linear Maxwell equations
\begin{equation}
\partial _{\alpha }\left( \frac{\partial V}{\partial F_{\alpha \beta }}%
\right) =J^{\beta }(x).  \label{EQ5}
\end{equation}
Here we have explicitly used current conservation $\partial _{\beta
}J^{\beta }=0,$ which is recovered from Eq.(\ref{EQ5}).
To make contact with the  formulation of Ref. \cite{PLEBANSKI} we identify the excitation tensor
$
P^{\alpha \beta }={\partial V}/{\partial F_{\alpha \beta }}
$, which contains the fields $\mathbf{D}$ and $\mathbf{H}$.
After the imposition of the Bianchi identity for the field strength in (\ref{ACT1}) we
recover the standard local action%
\begin{equation}
S(A_{\alpha })\;=\int d^{4}x\;\left( -\frac{1}{2} V(F_{\alpha \beta })- J^{\mu }A_{\mu
}\right), \label{ACFIN}
\end{equation}
which reproduces the equations of motion (\ref{EQ5}).
The most general form of the potential $V(F_{\mu\nu})$ will be an arbitrary function of the two invariants $F=F^{\mu\nu}F_{\mu\nu}$ and $G=\epsilon_{\alpha\beta\mu\nu}F^{\alpha\beta} F^{\mu\nu}$ \cite{EU}.
Let us observe that the choices $B=\frac{1}{2}F_{\mu\nu}\,dx^\mu \wedge dx^\nu; \,\, X=X_\mu\, dx^\mu; \,\, F=- {\rm d}\wedge\,X$ allows us to rewrite
(\ref{ACT1}) as a modified B-F action
\begin{equation}
S(B,F)= \int { B} \wedge { F} + S_{V({B})} + S_{J,{B}}.
\end{equation}
Using the above formulation we now consider the spontaneous symmetry breaking effective theory arising from the action
(\ref{ACT1}), which is obtained by expanding the fields around the non-zero
constant field configuration $C_{\mu\nu}$ which
minimizes the energy and maintains translational invariance. As it was  shown in Ref. \cite{JALU}, starting
from the energy-momentum tensor obtained from the action (\ref{ACT1}), such condition is
\begin{equation}
\left( \frac{\partial V}{\partial F^{\alpha \beta }}\right) _{C}=0.\label{MINCOND}
\end{equation}
Next we expand around the minimum setting%
\begin{equation}
F_{\alpha \beta }=C_{\alpha \beta }+f_{\alpha \beta },\;\;X_{\alpha
}=C_{\alpha }+\bar{X}_{\alpha },  \label{SHIFT}
\end{equation}%
which defines  the physical photon excitations $f_{\alpha \beta}$. The above expansion leads to the action
\begin{equation}
S=\frac{1}{2}\int d^{4}x\;\left( - \bar{V}-\epsilon ^{\nu \mu \alpha \beta }f_{\alpha \beta }\partial _{\nu }\bar{X}%
_{\mu }+f_{\mu \nu }J^{\mu \nu }\right) ,  \label{NBFACTION}
\end{equation}%
where we have dropped infinite constants together with total derivatives
(assuming that $\bar{X}_{\mu }$ goes to zero rapidly enough at infinity)
arising from the shift (\ref{SHIFT}). Here
\begin{equation}
\bar{V}=\bar{V}(f_{\alpha \beta })=V(C_{\alpha \beta }+f_{\alpha \beta
}). \label{FIFMC}
\end{equation}%
The equations of motion are the same
as in Eq. (\ref{EQ4}) with the replacements
$
V \rightarrow \bar{V}$, $F_{\alpha \beta }\rightarrow f_{\alpha \beta
}$, $X_{\alpha }\rightarrow \bar{X}_{\alpha}.
$
The Lagrange multipliers can be determined, up to a gauge transformation,
from the corresponding equations (\ref{EQ4}). Their elimination leads to  the final
Maxwell equations
\begin{equation}
\Sigma _{\alpha \beta \mu \nu }(f)\partial ^{\alpha }f^{\mu \nu }=J_{\beta },
\quad \epsilon _{\alpha \beta \mu \nu }\partial ^{\alpha }f^{\mu \nu }=0, \label{EMFIN}
\end{equation}%
where we have introduced the  full non-linear permeability $\Sigma _{\alpha \beta \mu
\nu }=\frac{\partial ^{2}\bar{V}}{\partial f^{\alpha \beta }\partial
f^{\mu \nu }}$. In the following we restrict ourselves to potentials such that $V=V(F)$  only.
In this case ${\bar V}={\bar V}(f_{\mu\nu}f^{\mu\nu} + 2 C_{\mu\nu}f^{\mu\nu} )$ with ${\bar V}'(f_{\mu\nu}=0) =0$ being the translation of
minimum condition  in Eq.(\ref{MINCOND}).The prime indicates a derivative with respect to the argument of  ${\bar V}$.
The expansion (\ref{FIFMC}) yields
\begin{equation}
\left(4 (C_{\alpha\beta}+f_{\alpha \beta}) (C_{\mu\nu}+f_{\mu \nu}) {\bar V}'' + 2 I_{\alpha\beta\mu\nu}{\bar V}'  \right)
\partial^\beta f^{\mu\nu}=0. \label{EEXPLICIT}
\end{equation}
for the first Eq. (\ref{EMFIN}). The notation is
$I_{\alpha\beta\mu\nu}=\eta_{\alpha\mu}\eta_{\beta\nu}-\eta_{\alpha\nu}\eta_{\beta\mu}$. The second of Eqs. (\ref{EMFIN})
implies that we can write $f_{\mu \nu}=\partial_\mu a_\nu- \partial_\nu a_\mu $.

It is  instructive to compare
the above Eq. (\ref{EEXPLICIT}) with the corresponding equation arising from a  so called bumblebee model, where the SLSB is
produced by a non-zero VEV $C^\mu$ of the potential $A^\mu$. In this case, starting from the Lagrangian
$L(A_\mu)=-F^{\mu\nu}F_{\mu\nu}/4- W(A^\mu A_\mu)$, with $W$  having a minimum
at $A^\mu=C^\mu$, and expanding $A^\mu=a^\mu + C^\mu$, the analogous equation results in
\begin{equation}
\partial_\mu f^{\mu\nu}- 2(a^\nu+ C^\nu) {\bar W}'=0. \label{EEXPLICITBB}
\end{equation}
Both Eqs.(\ref{EEXPLICIT}) and (\ref{EEXPLICITBB}) are non linear, mainly due to the presence of terms like ${\bar V}', {\bar V}'', {\bar W}'$. Since $V$ and $W$ have a minimum at $C_{\mu\nu}$ and $C_\mu$, the expansion of ${\bar V}'$ and ${\bar W}'$ in  terms of $f_{\mu\nu}$  and $a_\mu$  starts as  $(p\, C_{\mu\nu} f^{\mu\nu})$ and $( q \, C_\mu  a^\mu)$, respectively.
A question that naturally arises, which is relevant to study the propagation characteristics of the Goldstone modes, is to determine  the proper linearized approximation of these equations. In the case of Eq. (\ref{EEXPLICITBB}), it is clear that a correct  linear approximation is
$\partial_\mu f^{\mu\nu}-2q C_\mu C_\nu a^\nu =0$, which preserves the number of original degrees of freedom of the theory. In the alternative case of Eq. (\ref{EEXPLICIT}), a naive  analysis would suggest that
\begin{equation}
4 {\bar V}''(C) \, C_{\alpha\beta}C_{\mu\nu} \, \partial^\beta f^{\mu\nu} =0, \label{UNCLA}
\end{equation}
is the adequate linearized approximation. Nevertheless, this is not the case and one can verify that
Eq.(\ref{UNCLA}) describes negative degrees of freedom in four dimensions. The persistence of the term ${\bar V}' I_{\alpha\beta\mu\nu}$ in Eq. (\ref{EEXPLICIT}) is crucial to maintain the original two DOF arising from the action (\ref{ACT1}). In this way, Eq. (\ref{EEXPLICIT}) cannot be linearized around $f^{\mu\nu}=0$, but it is necessary to perform this operation around an additional field $f^{\mu\nu}_0$ which is solution of the full  non linear equation (\ref{EEXPLICIT}) and  such that ${\bar V}'(f^{\mu\nu}_0) \neq 0$.
Let us  observe that ${\bar V}\rightarrow {\bar V}(f_{\mu\nu}f^{\mu\nu})$,  in the limit $C_{\mu\nu} \rightarrow 0$. In this way, Eq. (\ref{EEXPLICIT}) does not provide small corrections in $C_{\mu\nu}$ to standard electrodynamics, but rather to an alternative non-linear theory.

\section{(III) The Goldstone theorem}
Our starting point is the effective potential $V(F)$, which we assume to
arise in an effective low energy theory that is obtained by integrating some
degrees of freedom up to certain scale in a more fundamental theory \cite{JALU}.  An
stable quantum theory is to be constructed around the minimum of the
potential given by $C_{\mu\nu}$. Since the symmetry is broken by the field
strength $F_{\alpha \beta }$, instead of the potential $A_{\mu }$, gauge
invariance is preserved from the outset. Usually the GM are identified as
zero mass excitations, but in this case gauge invariance guarantees that all
excitations are massless, which requires an alternative way of identifying
and interpreting such modes.

The corresponding incarnation of the Goldstone theorem is obtained by
following similar steps as in the case of the linear sigma model, for
example. We require the potential $V$ to be invariant under arbitrary
infinitesimal active  Lorentz transformations generated by  $G_{\;\;\alpha
}^{\mu }$,
\begin{equation}
\delta F^{\mu \nu }=G_{\;\;\alpha }^{\mu }F^{\alpha \nu }+G_{\;\;\alpha
}^{\nu }F^{\mu \alpha },  \label{GENLT}
\end{equation}%
where the invariance condition leads to
\begin{equation}
0=\delta F^{\mu \nu }\frac{\partial V}{\partial F^{\mu \nu }}.
\label{INVCOND}
\end{equation}%
Since this condition is valid for arbitrary $F^{\mu \nu }$, we can take an additional
derivative with respect to $F^{\alpha \beta }$ obtaining%
\begin{equation}
0=\frac{\partial \delta F^{\mu \nu }}{\partial F^{\alpha \beta }}\frac{%
\partial V}{\partial F^{\mu \nu }}+\delta F^{\mu \nu }\frac{\partial ^{2}V%
}{\partial F^{\alpha \beta }\partial F^{\mu \nu }}.  \label{SECONDDER}
\end{equation}%
Next we evaluate the above equation at the minimum of the potential,
which yields
\begin{equation}
0=\delta C^{\mu \nu }\left( \frac{\partial ^{2}V}{\partial F^{\alpha
\beta }\partial F^{\mu \nu }}\right) _{C}.  \label{TEOG1}
\end{equation}%
The previous equation splits into two cases according to the choice
of the vacuum $C^{\alpha \beta }$. There are two families of generators:
(i) those $\hat{G}_{\;\;}^{\alpha \beta }$ that leave the vacuum $C^{\mu \nu
} $ invariant ($\delta_{\hat G} C^{\mu \nu}=0$) and for which (\ref{TEOG1}) is
automatically satisfied and  (ii) those generators $\tilde{G}^{\alpha \beta }$,
which do not leave the vacuum invariant ( ${\delta_{\tilde G}}%
C^{\mu \nu }\neq 0$). In the latter case,  (\ref{TEOG1}) implies that the matrix
\begin{equation}
\Sigma _{\alpha \beta \mu \nu }^{(4)}\equiv\left( \frac{\partial ^{2}V}{\partial
F^{\alpha \beta }\partial F^{\mu \nu }}\right)_{C},
\end{equation}%
has zero modes  ${\delta_{\tilde G}}C^{\mu \nu }$.
These are precisely the GM of the theory.
\section{(IV) The choice of the vacuum and the identification of the Goldstone
modes}
The vacuum is given by  a constant electromagnetic tensor $C^{\mu
\nu }$ or equivalently, by a constant electric field $\mathbf{e}$ together
with a constant magnetic induction $\mathbf{b}$. Those two vectors determine
a plane, which can always be reached by adequate passive rotations of the
coordinate system. We still have the freedom to perform a passive Lorentz
boost in the direction perpendicular to the plane. At this stage, two
possibilities arise: (1) $\mathbf{e}$ and $\mathbf{b}$ are not orthogonal
and can be made parallel with this boost. (2) $\mathbf{e\;}$and $\mathbf{b\;}
$\ are orthogonal and will remain so after the boost. Choosing the plane of $%
\mathbf{e}$ and $\mathbf{b}$ as the $y-z$ plane we realize such two cases by taking
\begin{equation}
\mathbf{C}_1=\left[
\begin{array}{cccc}
0 & 0 & 0 & e \\
0 & 0 & b & 0 \\
0 & -b & 0 & 0 \\
e & 0 & 0 & 0%
\end{array}%
\right] ,\;\;\;\;\mathbf{C}_2=\left[
\begin{array}{cccc}
0 & 0 & e & 0 \\
0 & 0 & b & 0 \\
e & -b & 0 & 0 \\
0 & 0 & 0 & 0%
\end{array}%
\right],
\label{VACCHOICE}
\end{equation}%
respectively. The notation is $\mathbf{C}=\left[ C_{\;\;\nu }^{\mu }\right]$. This  is the most general parametrization of the
vacuum, which was introduced in Ref. \cite{KOSTELECKY5}.
Once we have chosen a vacuum, the general strategy to identify
the Goldstone modes  is presented in the following.
We start from an infinitesimal active Lorentz
transformation upon an antisymmetric tensor $H^{\mu \nu }$  which can be written in
matrix form as%
\begin{equation}
\delta \mathbf{H=}\left[ \mathbf{G,\;H}\right] ,\;\;\;\mathbf{H=}[ H_{%
\hspace{0.75em}\nu }^{\mu }].
\end{equation}%
Let us denote by $\mathbf{G}_{A}, \, A=1,\cdots ,6$,  a convenient choice of
the six independent Lorentz generators, which is adapted to a given vacuum
according to the following considerations. We separate the generators $%
\mathbf{G}_{A}$ into two groups: (i)\ the first one $\{ \mathbf{G}_{%
\tilde{A}},\;\;\tilde{A}=1,2,...,p\} $, which includes the linearly
independent generators that do not leave the vacuum invariant, i. e. $\left[ \mathbf{G}_{\tilde{A}},\mathbf{C}\right]\neq0$  and (ii) the
second one \ $\{ \mathbf{G}_{\hat{A}},\;\hat{A}=p+1,\;\ 6 \}$, %
such that $\left[ \mathbf{G}_{\hat{A}}, \mathbf{C}\right] =0$.
Consequently,  we split the indexes  $(A)$ into $({\tilde A}, {\hat A})$.
The GM $\mathbf{\Theta }^{\tilde{A}}$ are then identified as
those linearly independent tensors arising from the
transformations $\mathbf{G}_{\tilde{A}}$ acting on the vacuum, i.e.
$
\mathbf{\Theta }^{\tilde{A}}=\delta _{\tilde{A}}%
\mathbf{C=}\left[ \mathbf{G}_{\tilde{A}},\;\mathbf{C}\right].
$
The linear independence of the $\mathbf{\Theta
}^{\tilde{A}}$ is guaranteed by that of the generators $\mathbf{G}_{%
\tilde{A}}$. It is clear that any linear combination of $\delta _{\tilde{A}}%
\mathbf{C}$ is also a GM. A basis $\{ {\mathbf \Theta}^A \}$  for the antisymmetric tensor space is then completed by
conveniently choosing the remaining linearly independent modes $\mathbf{%
\Theta }^{\hat{A}}$.
We find it convenient to parameterize the action of the Lorentz group upon the vacuum as
\begin{equation}
\delta_{A} \mathbf{C}=[\mathbf{G}_A, \mathbf{C}]= R_{(A)(B)}\,{\mathbf \Theta}^B,
\label{ACTVAC}
\end{equation}
where $R_{{(\hat A)} (B)}=0 $.

A very compact notation can be achieved by projecting antisymmetric tensor
indices $K_{....\mu \nu ....}$ into the basis $\mathbf{\Theta }^{A}=(
\mathbf{\Theta }^{\tilde{A}},\;\mathbf{\Theta }^{\hat{A}}) $, by
defining
$
K...^{(A)}.....=\Theta ^{(A)\mu \nu }K_{....\mu \nu ....}
$
for each antisymmetric pair of indices $\mu \nu $.

An important information is the
transformation properties of the chosen basis $\;\mathbf{\Theta }^{(A)}\;$%
under the associated Lorentz transformations. Clearly, the transformations $%
\delta _{B}\mathbf{\Theta }^{(A)}$ are such that
\begin{equation}
\delta _{B}\mathbf{\Theta }^{(A)}=\left[ \mathbf{G}_{B},\;\mathbf{\Theta }%
^{(A)}\right] \equiv C_{B\;\;\;\;(M)}^{\;\;(A)}\;\mathbf{\Theta }^{(M)},
\label{COEFFLT}
\end{equation}%
where $C_{B\;\;\;\;(M)}^{\;\;(A)}$ is the corresponding representation of the infinitesimal Lorentz transformation induced by the generator $%
\mathbf{G}_{B}$. Given an specific choice of a basis\ $\mathbf{\Theta }^{(A)}$,
the coefficients $C_{B\;\;\;\;(M)}^{\;\;(A)}$ can be readily determined. It will prove convenient to make use of the $SU(2)\times SU(2)$ decomposition of the Lorentz group. In terms of the standard generators $\mathbf{L}_a, \mathbf{K}_a, \,a=1,2,3$ we define $\mathbf{G}_{(\alpha a)}=\mathbf{L}_a+\alpha \mathbf{K}_a, \, \alpha= \pm$, which have the following useful properties
\begin{equation}
\left[ \mathbf{G}_{(\alpha a)},\mathbf{G}_{(\beta b)}\right] =2i\delta _{\alpha \beta
}\epsilon _{abc}\mathbf{G}_{(\beta c)}, \label{SU21}
\nonumber
\end{equation}
\begin{equation} \quad {\rm Tr}\left( \mathbf{G}_{(\alpha a)}\mathbf{G}_{(\beta
b)}\right)=4\delta _{\alpha \beta }\delta _{ab}. \label{SU2SU2}
\end{equation}
There is no summation when a repeated index in one side of the equation
appears as a free index in the other side.

Next we make explicit the above general strategy in
the case of the vacuum $\mathbf{C}_1$, defined in Eq.(\ref{VACCHOICE}. Using the representation of the Lorentz generators given in Eqs.(\ref{SU21}),(\ref{SU2SU2})
we rewrite
\begin{equation}
\mathbf{C}=-\frac{i}{2}\omega _{+}\mathbf{G}%
_{+3}-\frac{i}{2}\omega _{-}\mathbf{G}_{-3},\qquad \omega _{\alpha
}=-b+i\alpha e.
\end{equation}%
From the commutation relations (\ref{SU21}) it is clear that we can choose $%
\mathbf{G}_{+3},\;\mathbf{G}_{-3}$ as the linearly independent generators of
the invariance subgroup of the vacuum, which is isomorphic to $T(2)$. This
means that we have the identification $\{ \hat{A}\} =\left\{
+3,-3\right\} =\left\{ \alpha 3,\;\alpha =\pm \;\right\} $ and that we have
chosen $\mathbf{G}_{\hat{A}}:\mathbf{G}_{\alpha 3}$.
The remaining four generators
$
\mathbf{G}_{\tilde{A}}:\mathbf{G}_{\alpha \tilde{a}}\;,\;\tilde{a}%
=1,2, \,  \{\tilde{A}\} =\{ \alpha \tilde{a}\}
$
give rise to four GM.  In this case it is  convenient to define the GM as the action of a linear combination
of generators acting upon the vacuum, according to
\begin{eqnarray}
\mathbf{\Theta }^{(\alpha \tilde{a})}=\epsilon _{\tilde{a}\tilde{b}}\left[
\mathbf{G}_{\alpha \tilde{b}},\mathbf{C}\right], \quad \epsilon_{\tilde{1}\tilde{2}}=+1, \label{GM1}
\end{eqnarray}%
in such a way that they reduce to
\begin{equation}
\mathbf{\Theta }^{(\beta \tilde{b})}=\omega _{\beta \tilde{b}}\mathbf{G}%
_{\beta \tilde{b}},\qquad \omega _{\beta \tilde{b}}=\omega _{\beta }, \nonumber
\end{equation}
\begin{equation}\omega _{\beta 1}=\omega _{\beta 2}=( -b+i\beta e) \equiv \omega
_{\beta }\;.\;\;
\end{equation}%
The remaining two linearly independent members of the basis $\mathbf{\Theta }%
^{(A)}$ are chosen to be
$
\Theta ^{(\gamma 3)}=\omega _{\gamma }G_{\gamma 3}
$,
where the factor $\omega _{\gamma }$ has been inserted for dimensional
reasons and in order to simplify subsequent calculations. Finally, the basis for
the antisymmetric tensor space, which include the GM, can be written as
\begin{equation}
\mathbf{\Theta }^{(\alpha a)}=\omega _{\alpha a }\mathbf{G}_{\alpha a},\qquad
\omega _{\alpha 1}=\omega _{\alpha 2}=\omega _{\alpha 3}= ( -b+i\alpha e). \label{OMEGA}
\end{equation}%
From Eq. (\ref{GM1}) we identify the remaining coefficients $R_{(%
\tilde{A})(B)}$ defined in (\ref{ACTVAC}) as
\begin{equation}
\;\;\;\;R_{(\alpha \tilde{a})(\beta \tilde{b})}=-\delta _{\alpha \beta
}\epsilon _{\tilde{a}\tilde{b}}\;\;,\;\;\;R_{(\alpha \tilde{a})(\beta 3)}=0.\label{ACTVAC1}
\end{equation}
Since the Goldstone modes, living in the subspace spanned by $\{\Theta^{{\tilde A}}{}_{\mu\nu}, \, {\tilde A}= (\alpha, {\tilde a})=1,2,3,4 \}$ , comprise two independent electric excitations plus two independent magnetic excitations in phase space, they describe  at most two DOF. On the other hand, the two non-GM modes belonging to  the subspace with basis $\{\Theta^{{\hat A}}{}_{\mu\nu}, \, {\hat A}= (\alpha, 3)=1,2 \}$ can support at most one DOF.

The coefficients $C_{B\;\;\;\;(M)}^{\;\;(A)}$ defined in
(\ref{COEFFLT}), turn out to be
\begin{equation}
C_{\alpha a\;\;\;\;\;\;\left( \gamma c\right) }^{\;\;\;\;(\beta
b)}=2i\epsilon _{abc}\delta _{\alpha
\beta }\delta _{\beta \gamma },  \label{COEFFARB}
\end{equation}%
in the modified notation.

The case of the vacuum $\mathbf{C}_2$ in Eq. (\ref{VACCHOICE}) can be dealt with in a similar manner.
Again, there are four GM in phase space which account at most  for two DOF, together with two non-GM in phase space which can support at most one DOF.
\section{(VI) Relations among the permeabilities}
As an extension of the procedure used to derive the Goldstone theorem in Section (III), let
us observe that by taking additional derivatives with respect to the field
strength on the relation (\ref{SECONDDER}) and evaluating them in the
vacuum, we obtain relations among the generalized permeabilities $\;\Sigma
_{\alpha _{1}\beta _{1}\alpha _{2}\beta _{2}.......\alpha _{n}\beta
_{n}}^{(2n)}$ defined by
\begin{equation}
\Sigma _{\alpha _{1}\beta _{1}\alpha _{2}\beta _{2}.......\alpha _{n}\beta
_{n}.}^{(2n)}=\left( \frac{\partial ^{n}V}{\partial F^{\alpha _{1}\beta
_{1}}\partial F^{\alpha _{2}\beta _{2}}........\partial F^{\alpha _{n}\beta
_{n}}}\right) _{C}.
\end{equation}%
Such relations might be relevant by recalling that the expanded potential $%
\bar{V}(f)$, together with the full permeability $\Sigma _{\alpha \beta \mu
\nu }(f)$, can be written in terms of them as%
\begin{equation}
 \bar{V}(f_{\mu \nu
})= \sum_{n=2} \frac{1}{n!\times 2^{n}}\Sigma_{\alpha _{1}\beta _{1} \cdots\alpha _{n}\beta _{n}}^{(2n)}f^{\alpha _{1}\beta
_{1}}\cdots f^{\alpha _{n}\beta _{n}}
\label{EXP1}, \nonumber \\
\end{equation}
\begin{equation}
\hskip-.2cm \Sigma_{\alpha \beta \mu \nu }=\sum_{n=2}\frac{1}{2^{n-2}\left( n-2\right) !}\Sigma _{\alpha \beta
\mu \nu \alpha _{3}\beta _{3}\dots \alpha _{n}\beta
_{n}}^{(2n)}f^{\alpha _{3}\beta _{3}}
\dots f^{\alpha _{n}\beta _{n}},\nonumber \\
\label{EXP2}
\end{equation}%
respectively.

In fact, taking one additional derivative with respect to the field strength in (%
\ref{SECONDDER}) and evaluating the resulting equation at the minimum leads to
\begin{eqnarray}
&&(G_{A}){}^{\mu }{}_{\alpha }\Sigma _{\mu \beta \kappa \lambda
}^{(4)}+(G_{A}){}^{\mu }{}_{\beta }\Sigma _{\alpha \mu \kappa \lambda
}^{(4)}+(G_{A}){}^{\mu }{}_{\kappa }\Sigma _{\alpha \beta \mu \lambda }^{(4)} \nonumber \\
&&+(G_{A}){}^{\mu }{}_{\lambda }\Sigma _{\alpha \beta \kappa \mu }^{(4)}+%
\frac{1}{2}\left(\delta_{A}C^{\mu \nu }\right)\Sigma _{\mu \nu \alpha \beta \kappa
\lambda }^{(6)}=0.  \label{REL6TO4}
\end{eqnarray}%
Again, the relation (\ref{REL6TO4}) naturally splits into two cases
according to the choices $R_{(\hat{A})(B)}=0$ and $R_{(\tilde{A})(B)}\neq 0$.
Let us recall that the coefficients $R_{(A)(B)}$ have been already defined in Eq. (\ref{ACTVAC}) and previously evaluated in Eq. (\ref{ACTVAC1}) for the vacuum ${\mathbf C}_1$.
The projection into the basis $\mathbf{\Theta }^{(A)}$ produces%
\begin{eqnarray}
&&C_{\hat {A}\;\;\;(M)}^{\;(B)}\;\Sigma ^{(4)(M)(C)}+C_{\hat {A}%
\;\;\;(M)}^{\;(C)}\;\Sigma ^{(4)(B)(M)}=0,  \label{REL6TO41} \\
&&-\frac{1}{2}R_{(\tilde{A})(M)}\Sigma ^{(6)(M)(\tilde{B})(\hat{C})}=C_{%
\tilde{A}\;\;\;\left( \hat{M}\right) }^{\;(\tilde{B})}\;\Sigma ^{(4)(\hat{M}%
)(\hat{C})},\nonumber \\
\label{REL6TO42}
\end{eqnarray}%
respectively. The first of the above equations states that the components $
\Sigma ^{(4)(A)(B)}$ are an invariant tensor under the symmetry subgroup of
the vacuum. The second one provides a relation between the components of $%
\Sigma ^{(6)(\tilde{A})(B)(C)}$ and those of the lower order  permeabilities $%
\Sigma ^{(4)(\hat{M})(C)}$.
Taking further additional  derivatives in (\ref{SECONDDER}) and evaluating the result  at the minimum
we can generalize the above statements obtaining
\begin{eqnarray}
&& \delta _{\hat{A}}\Sigma ^{(2n)(B_{1})(B_{2})...(B_{n})}=0,  \label{GEN2N}
\\
&& \delta _{\tilde{A}}\Sigma ^{2n(B_{1})(B_{2})...(B_{n})}=-\frac{1}{2}R_{(%
\tilde{A})(M)}\times \nonumber \\
&&\Sigma ^{2(n+1)(M)(B_{1})(B_{2})...(B_{n})}.  \label{GEN2NP2}
\end{eqnarray}%
The formulation of the model in terms of the expansion (\ref{EXP1}) can be useful in the case where the potential $V(F)$ is not known
and one is dealing with a bottom-up approach.
\section{(VII) Degrees of freedom (DOF) and propagating Goldstone modes (GM)}
Since we have respected
gauge invariance, the two DOF that arise in the model  have to be massless.
The propagation properties of NLED are normally characterized by selecting a
given background $f_{0}^{\mu \nu }$, solution of the non-linear equations,
and studying perturbations around it, either via the behavior of field
discontinuities across a given surface, which are determined by the Fresnel
equations \cite{PLEBANSKI}, or just by an appropriate linearization of Eq. (\ref{EEXPLICIT}): $f^{\mu\nu}=f_0^{\mu\nu} +(f_L)^{\mu\nu}$. In our case, there is the further condition
that ${\bar V}'(f_{0}^{\mu \nu }) \neq 0$.
The field $(f_L)^{\mu\nu}$ will characterize the DOF of the model.
On the other hand, the GM were defined in Section (IV) as certain linear combinations of electric and magnetic
fields which are null tensors of the fourth order permeability $\Sigma
^{(4)\alpha \beta \mu \nu }$. Let us  emphasize again that the GM in our case,  are not at all related to any mass matrix.
The propagation properties of the GM, which will depend both upon the vacuum $C^{\mu\nu}$ and
the selected linearization point $f_0^{\mu\nu}$,  can be subsequently obtained by projecting them from the solution $(f_L)^{\mu\nu}$.
In fact, as discussed in section (IV), the basis for the antisymmetric tensor space $\Theta^{A}{}_{\mu\nu}$ can be always divided into two
orthogonal subsets: (i) $\{\Theta^{{\hat A}}{}_{\mu\nu}, \, {\hat A}=1,2  \}$ arising from the generators that leave the vacuum invariant and (ii)
$\{\Theta^{{\tilde A}}{}_{\mu\nu}, \, {\tilde A}=1,2,3,4, \, \Theta^{{\tilde A}}{}^{\alpha\beta}\Theta^{{\tilde B}}{}_{\alpha\beta}=N_{{\tilde A}}\delta^{{\tilde A}{\tilde B}}  \}$, which span the subspace containing the GM. Then, the contribution of the GM to the propagating DOF of the model is given by
\begin{equation}
f^{(GM)}_{\mu\nu}=\sum_{{\tilde A}}\frac{1}{N^{{\tilde A}}}\left( \Theta^{({\tilde A})}{}_{\alpha\beta}(f_{L})^{\alpha\beta}\right)  \Theta^{{\tilde A}}_{\mu\nu}.
\end{equation}
The non-GM are given by a similar projection of $(f_{L})^{\mu\nu}$ onto the subspace generated by $\Theta^{{\hat A}}{}_{\mu\nu}$. It may well happen that $(f_{L})^{\mu\nu}$ has non-zero projection in both subspaces. In this way, the splitting of the DOF into GM and non-GM has to be decided in each particular case.
The only general statement that we can make is that the GM can account at most for two DOF, while the non-GM can account  at most for one DOF.
\section{(VI) Summary and Conclusions}
In this work we study the spontaneous Lorentz symmetry breaking arising in
non-linear electrodynamics (NLED), induced by a constant non-zero vacuum expectation value
$C_{\mu \nu }$ of
the field strength $F_{\mu \nu }$.
In this way
gauge invariance is maintained from the outset and all excitations are massless. The main point we deal with
is the reinterpretation and characterization of the Goldstone theorem in
this situation. Since our basic field is $F_{\mu \nu }$ we consider a first order
formulation of  NLED, which is
different from the  standard Plebanski formulation and has some advantages
in this case: the excitation fields ($\mathbf{D},\mathbf{H}$ unified in $%
P_{\mu \nu }$) are directly given in terms of the field strengths ($\mathbf{E%
},\mathbf{B}$ unified in $F_{\mu \nu }$), without the need of inverting the
non-linear relation $F_{\mu \nu }=F_{\mu \nu }(P_{\alpha\beta})$. We start from an action
including a Lorentz invariant
potential $V(F_{\mu \nu })$, which we assume to arise from a fundamental
theory, having an absolute  minimum at $F_{\mu \nu }=C_{\mu \nu }$, that defines the
vacuum of the model. With this potential, the action $S(F_{\mu \nu
},\;X_{\alpha })$ in Eq.(\ref{ACT1}) is constructed, which includes Lagrange multipliers $%
X_{\alpha }$  required to recover the Bianchi identity for the field
strength in the final NLED.  This action describes two DOF. Under the conditions
of constant field configurations, the vacuum of the model is given by  the minimum of $V$.
Expanding  the fields around the minimum defines the physical excitations
contained in $f_{\mu \nu }=F_{\mu \nu }-C_{\mu \nu }$.
Further elimination of the Lagrange multipliers leads to a NLED defined by a
Lagrangian $L(a_{\mu })=-\bar{V}(f_{\mu \nu }=\partial _{\mu }a_{\nu
}-\partial _{\nu }a_{\mu })/2-J_{\mu }a^{\mu }$. The expansion of $%
\bar{V}$ in powers of $f_{\mu \nu }$ allows for the introduction of
generalized permeabilities $\Sigma _{\alpha _{1}\beta _{1}\alpha
_{2}\beta _{2}.......\alpha _{n}\beta _{n}}^{(2n)}$ in a manner similar to
when the polarization is expanded in terms of
the susceptibilities when studying non-linear effects in standard optics
\cite{NLO}. In the limit of small SLSB ($C_{\mu\nu} \rightarrow 0$) our model does not reduce
to standard electrodynamics plus small corrections, but instead  produces corrections to an alternative NLED.
Only after a further linearization $f^{\mu \nu }=f^{\mu \nu }_0 + (f_L)^{\mu \nu }$, where $f^{\mu \nu }_0$ is some exact solution of Eqs.(\ref{EEXPLICIT}), we obtain terms proportional to
$ (f_L)_{\mu \nu }(f_L)^{\mu \nu}$ in the Lagrangian, thus making contact with the photon sector of the SME.

The imposition of active  Lorentz invariance upon the potential $V$,
together with  further iterations of the procedure leading to the Goldstone theorem, yields
the following properties of the permeabilities: (1) \ $\Sigma _{\alpha
\beta}^{(2)}=0$, which is just a consequence of the minimum condition. (2)
$\ \Sigma _{\alpha \beta \mu \nu}^{(4)}$ has zero modes $\tilde{\Theta}^{\mu
\nu }$, which are the Goldstone modes of the model, that are generated by the infinitesimal active Lorentz transformations which do not leave the vacuum invariant. This constitutes the statement of
the Goldstone theorem in our case. (3)
the permeabilities $\Sigma _{\alpha _{1}\beta _{1}
.......\alpha _{n}\beta _{n}}^{(2n)}$ are invariant tensors under the transformations that
leave the vacuum invariant. (4) those transformations that do not leave the
vacuum invariant impose relations among the permeabilities of order  $2n$ and order $2n+2$.

Next we identify the Goldstone modes. In the case of the Lorentz group in
four dimensions, only two inequivalent vacua exist. Choosing the $SU(2)\times SU(2) $ representation of the Lorentz generators, both cases can be deal with in analogous form, provided $e^2\neq b^2$.
We have verified that for any choice of
the parameters $e$ and $b$, which define the vacuum, there are always
four  generators $G_{\tilde{A}}$ that do not leave the vacuum invariant, plus two generators
$G_{\hat{A}}$ that leave the vacuum invariant. That is to say, the invariant subgroup of the model is always isomorphic to the translation group in two dimensions $T(2)$. The partition of the generators implies that  the basis for the antisymmetric tensor space can be split into two
orthogonal sets $\{\Theta^{\tilde A}{}_{\mu\nu}\}$ and  $\{\Theta^{\hat A}{}_{\mu\nu}\}$ in phase space, which span
the GM  and the non-GM respectively. This means that
the GM can support at most two DOF while the non-GM can account at most for one DOF.

The propagating properties of the DOF of the model are dictated by  $(f_L)^{\mu\nu}$, which is obtained  as a solution of the linearized
equations of motion. In our case, the kinetic term
of the equations of motion depends upon the linearization process and one has to be careful in keeping it
in order to maintain the initial two DOF. This is a main difference with a standard bumblebee model, where the kinetic term  is independent of
the approximation considered in the potential. The propagation properties of the GM are subsequently obtained by projecting $(f_{L})^{\mu\nu}$ in the corresponding subspace. A detailed description of such properties is beyond the scope of the present letter and it will be discussed elsewhere.
\acknowledgments
L.F.U is partially supported by the projects DGAPA-IN111210, DGPA-IN109013, together with a sabbatical fellowship from DGAPA-UNAM. He also acknowledges the hospitality of the Facultad de F\'\i sica, together with support from the Programa de Profesores Visitantes, at the Pontificia Universidad Cat\'olica de Chile. C.A.E acknowledges support from a CONACyT graduate fellowship as well as partial support from the project DGAPA-UNAM-IN109013 and the program PAEP at UNAM.

\end{document}